\documentstyle[epsfig,sprocl]{article}
\def\Journal#1#2#3#4{{#1} {\bf #2}, #3 (#4)}

\def\NIMA{{\em Nucl. Instrum. Methods} A}

\def\NPBP{{\em Nucl. Phys.} B (Proc. Suppl.)}

\def\ra{\rightarrow}
\def\be{\begin{equation}}  
\def\ee{\end{equation}}
\def\bea{\begin{eqnarray}}
\def\eea{\end{eqnarray}}
\newcommand{\lsnd}{LSND } 
\newcommand{\nue}{neutrino } 
\newcommand{\nues}{neutrinos }

\newcommand{\noszp}{neutrino oscillations. }
\newcommand{\osz}{oscillation }
\newcommand{\oszsp}{oscillations. }
\newcommand{\oszs}{oscillations }

\newcommand{\sk}{Superkamiokande }

\newcommand{\delm}{\mbox{$\Delta m^2$} }

\newcommand{\nel}{\mbox{$\nu_e$} }
\newcommand{\bnel}{\mbox{$ \bar \nu_e$} }
\newcommand{\bnmu}{\mbox{$ \bar \nu_{\mu}$} }
\newcommand{\nmu}{\mbox{$\nu_\mu$} }
\newcommand{\ntau}{\mbox{$\nu_\tau$} }
\newcommand{\sint}{\mbox{$sin^2 2\theta$} }
\newcommand{\lbls}{long baseline experiments }

\begin{document}
\title{STATUS OF NEUTRINO OSCILLATION SEARCHES  \footnote{to appear in Proc. COSMO'97, Ambleside (UK), September
1997}}
\author{ K. ZUBER}
\address{Lehrstuhl f\"ur Experimentelle Physik IV, Universit\"at Dortmund,\\
44287 Dortmund, Germany}
\maketitle
\abstracts{The current status of \nue \osz searches with reactors and accelerators is reviewed.
An outlook, especially
on future long baseline \nue \osz projects, is given.}
\section{Introduction}
The existence of massive \nues opens up a variety of new phenomena which could
be investigated by experiments. One of these is \nue \oszsp 
In the simplified picture of two flavour \oszs they can be parametrized by two parameters, \sint and $\Delta
m^2$. 
While $\sint$ describes the amplitude of the
oscillation, $\delm = m_2^2 -m_1^2$ determines the oscillation length given in practical units as
\be
L = \frac{4\pi E \hbar}{\Delta m^2 c^3} =
2.48 (\frac{E}{MeV})(\frac{eV^2}{\Delta m^2}) \quad m
\ee
As can be seen, \oszs do not allow an absolute mass measurement and \nues must not be exactly degenerated. 
For a discussion of direct bounds on \nue masses see \cite{cald}. 
From first principles, there is no preferred region
in the $\delm - \sint$
parameter space and therefore the whole has to be investigated experimentally.\\
On earth, two artificial \nue sources exist in form of nuclear power reactors and accelerators.
For a more detailed overview see \cite{hd}.
\section{Reactor experiments}
\begin{figure}[hhh]
\begin{center}
\epsfig{file=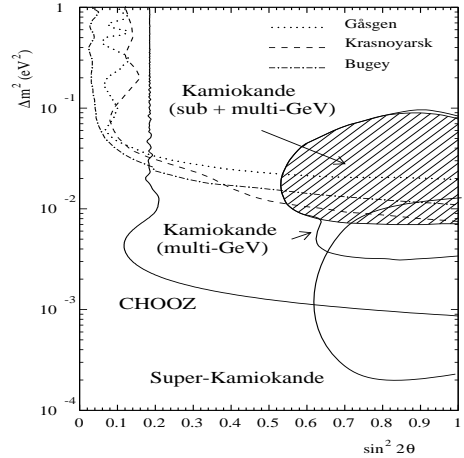,width=6cm,height=6cm}
\end{center}
\caption{\label{chores} \it Exlcusion plot for \bnel - $\bar{\nu}_X$ \osz as given by the
CHOOZ-results and other reactor experiments. Also shown are the 
allowed regions from atmospheric \nues . As can be seen, all of the Kamiokande and most of the 
Superkamiokande region
is excluded and therefore support the \nmu - \ntau \osz scenario as an explanation for the atmospheric
\nue deficit.}
\end{figure}

 Reactor experiments are looking for
\bnel $\ra \bar {\nu}_X$ disappearance.
Reactors are a source of MeV \bnel, due to the fission of nuclear fuel.
Experiments typically 
try to measure the positron spectrum which can be deduced from the \bnel - spectrum
and either compare it directly to the theoretical predictions
or measure it at several distances from the reactor and search for spectral distortions. Both
types of experiments were done in the past. The detection relies on the reaction
\be
\label{gl1}
\bnel + p \ra e^+ + n
\ee
with an energy threshold of 1.804 MeV.
Different strategies are used for the detection of the
positron and the neutron. Normally, coincidence techniques are used between the annihilation photons and the
neutrons 
which diffuse and thermalise within 10-100 $\mu$s and materials like Gd are then used 
for neutron-capture.
The most recent experiment is CHOOZ in France \cite{chooz}.
Compared to previous experiments, this detector has some advantages.
First of all, the detector is located underground with
a shielding of 300 mwe, reducing the background due to cosmics by a factor of 300. Moreover, 
the detector is about 1030 m away
from the reactor (more than a factor 4 in comparison to previous experiments)
enlarging the sensitivity to smaller $\Delta m^2$. In addition,
the main target has about 4.8 t of a specially developed Gd-loaded scintillator
and is therefore much larger than those used before. First results can be seen in Fig.\ref{chores}.\\
An upcoming experiment is the Palo Verde (former San Onofre) experiment \cite{pave} near
Phoenix, AZ (USA). It will consist of a 12 t liquid scintillator
also loaded with Gd. 
The experiment will be located under a shielding of 46 mwe in a distance
of about 750 (820) m to the reactors. The experiment will be online by
late 1997.\\
A first long-baseline 
reactor experiment (KamLAND) \cite{kamland} using a 1000 t liquid scintillator detector at 
the Kamioka site in a distance
of 150 km to a reactor is approved by the Japanese Government. It could start data taking in 2000.
\section{Accelerators}
Accelerators typically produce \nue beams by shooting a proton beam on a fixed target.
The produced secondary pions and kaons decay and create a
\nue beam dominantly consisting of $\nu_{\mu}$.
The detection relies on charged current reactions $\nu_i N \ra i + X  \quad (i= e, \mu, \tau$),
where N is a nucleon and X the hadronic final state.
Depending on the intended goal, the search for
\oszs therefore requires a detector
which is capable of detecting electrons, muons and $\tau$ - leptons in the final
state.
Accelerator experiments are mostly of appearance type working in
the channels \nmu - $\nu_X$ and \nel - $\nu_X$.
\subsection{Accelerators at medium energy}
\begin{figure}[hhh]
\begin{center}
\begin{tabular}{cc}
\epsfig{file=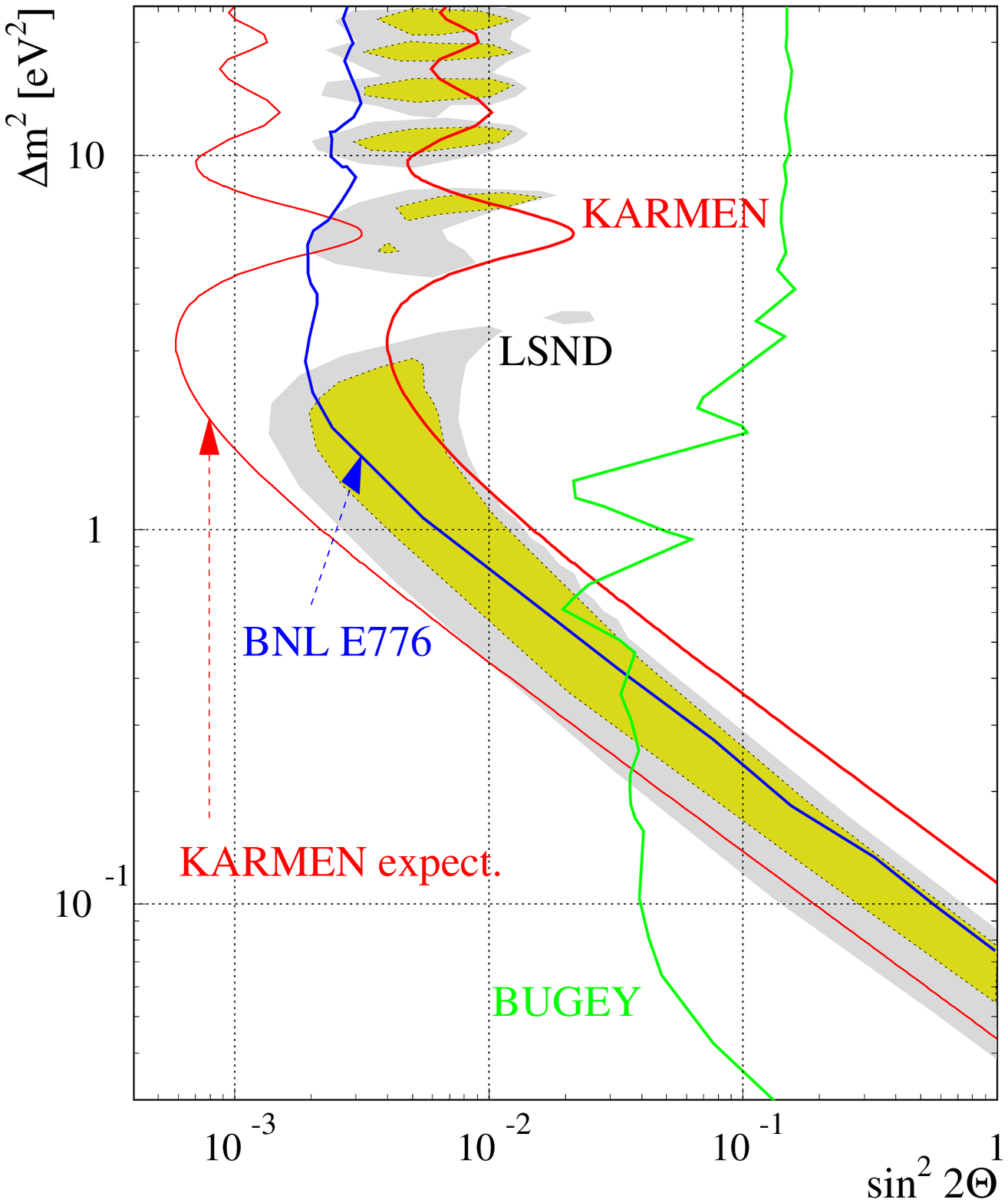,width=6cm,height=6cm} &
\epsfig{file=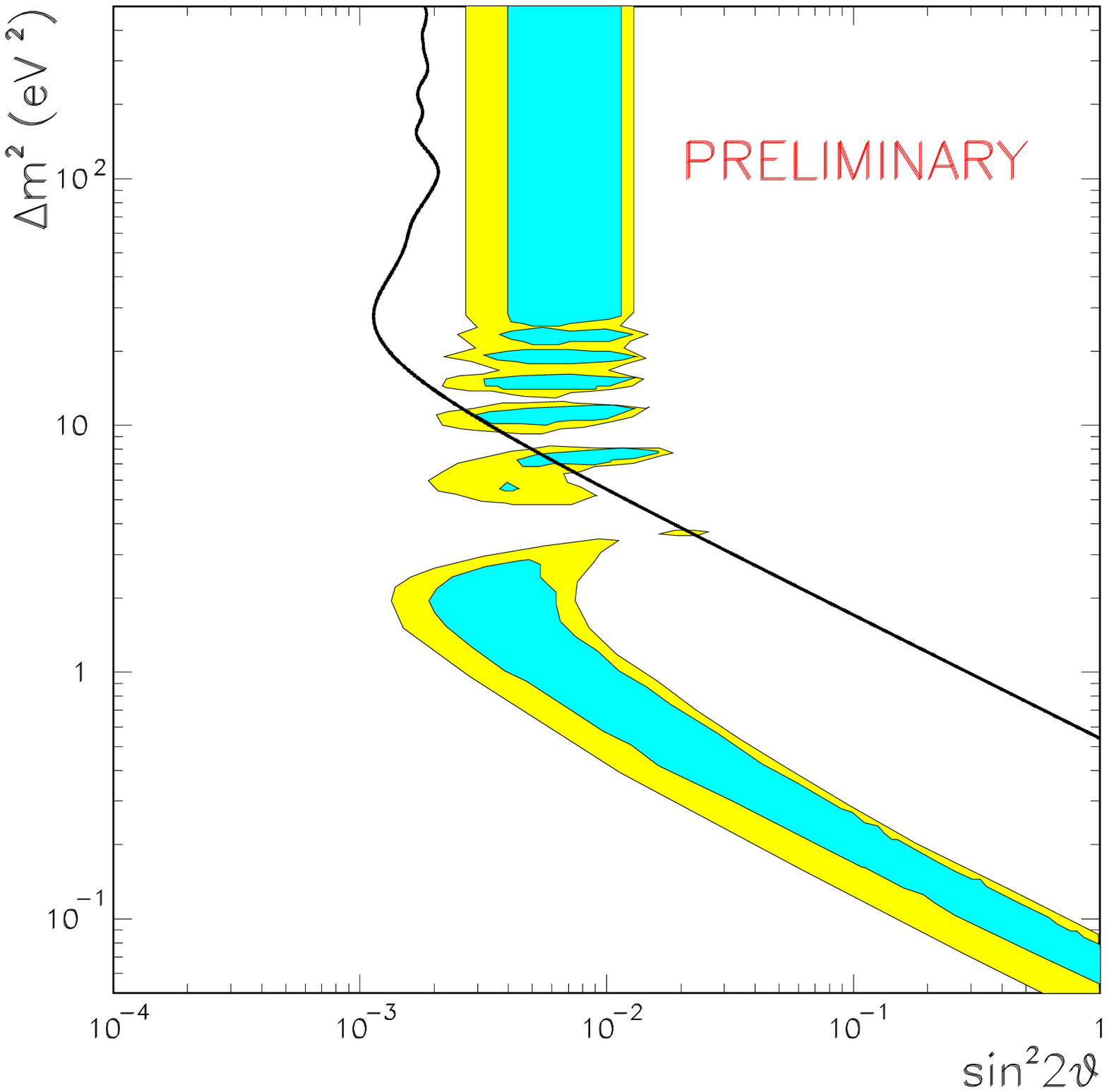,width=6cm,height=6cm}
\end{tabular}
\end{center}
\caption{\label{lsndev} \it Region of evidence for \bnmu - \bnel \oszs from \lsnd together with already
excluded
parts from KARMEN, E776 and the Bugey reactor experiment. Also shown is the possibility of KARMEN after running
three more years with the new upgrade. The second picture shows the region excluded by NOMAD.}
\end{figure}

At present there are two experiments running with \nues at medium energies
($E_\nu \approx $ 30 - 50 MeV) namely KARMEN \cite{karmen} and \lsnd \cite{lsnd}.
The limits reached so far and the \lsnd evidence are shown in Fig.\ref{lsndev}. 
Recently \lsnd  published their \nel - \nmu analysis for pion decays in flight which is in
agreement with the former evidence from pion decay at rest \cite{lsnd}.
To improve the sensitivity by reducing the 
neutron background KARMEN constructed a
veto shield against atmospheric muons
which has been in operation since Feb.1997 and is surrounding the whole detector. 
The region which can be excluded in 2-3 years of running in the upgraded version is also shown in Fig.\ref{lsndev}.
Also \lsnd continues with data aquisition.
\newline
To test the \lsnd region of evidence several new projects are planned.
The Fermilab 8 GeV proton booster offers the chance for a \nue experiment (BooNE) \cite{boone} which
could start data
taking in 2001.
An increase in sensitivity in the \nel - \nmu \osz channel could also be reached
by a proposed experiment at the CERN PS \cite{cernps} or if there is a possibility for
\nue physics
at the planned European Spallation Source (ESS)
or the National Spallation Neutron Source (NSNS) at Oak Ridge which might have a 1 GeV proton beam in 2004.
\subsection{Accelerators at high energy}
High energy accelerators provide \nue beams with an average energy in the GeV region.
Here, at present especially CHORUS and
NOMAD at CERN are providing new limits \cite{chorus}. 
Both experiments are 823 m (CHORUS) and 835 m (NOMAD) away from a beam dump and designed to
improve the existing limits on \nmu - \ntau \oszs by an order of magnitude. 
The present limits (Fig.\ref{sum}) for large \delm are \cite{chorus}
\bea
\sint < 2.1 \times 10^{-3} \quad (90 \% CL) \quad (CHORUS)\\
\sint < 3.4 \times 10^{-3} \quad (90 \% CL) \quad (NOMAD)
\eea
The final goal is to reach a sensitivity down to \sint $\approx 2 \times 10^{-4}$ for large $\Delta m^2$.
Having a good electron identification NOMAD also offers 
the possibility to search in the \nel - \nmu channel.
The exclusion plot is shown in Fig.\ref{lsndev} with a limit of $\sint < 2 \times 10^{-3} \quad (90 \% CL)$
for large \delm. While the CHORUS data taking is complete, NOMAD will continue 1998.
\section{Future accelerator experiments}
Possible future ideas split into two groups depending on the physical goal. One part is focussing on
improving the
existing bounds in the eV-region 
by another order of magnitude with respect to CHORUS and NOMAD and to investigate the \lsnd evidence.
Other groups plan to increase
the source - detector distance to probe smaller \delm and to be directly comparable to atmospheric scales.
\subsection{Short and medium baseline experiments} 
\begin{figure}[hhh]
\begin{center}
\epsfig{file=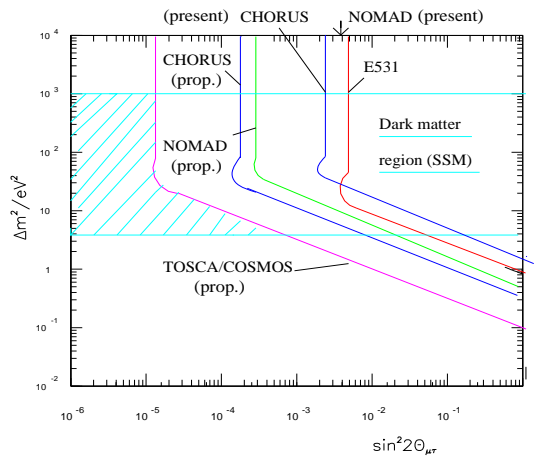,width=7cm,height=6cm}
\end{center}
\caption{\label{sum} \it \nmu - \ntau exclusion plot showing the present limits of CHORUS and NOMAD as well
as the
proposed limits of TOSCA and COSMOS. The shaded region corresponds to the region of an \ntau
in the eV-region motivated by dark
matter considerations and the quadratic see-saw-mechanism (SSM).}
\end{figure}

Several ideas exist for a next generation of short baseline experiments. At CERN the follow up could be
TOSCA \cite{tosca},
combining features of NOMAD and CHORUS. The idea is to use 2.4 tons of emulsions together with
large silicon microstrip detectors within the NOMAD magnet.
For TOSCA the option to extract a \nue beam at lower proton energies (350 GeV) at the CERN SPS exist.
At Fermilab the COSMOS (E803) experiment \cite{cosmos} at the new Main Injector is proposed. 
It will produce a
proton beam of 120 GeV resulting
in an average \nue energy of $\langle E_{\nu} \rangle \approx 12$ GeV.
The main target consists of emulsions with a total mass of 865 kg.
The distance to the beam dump will be 960 m.
Both experiments are designed to improve
the sensitivity in the \nmu - \ntau channel by
one order of magnitude (Fig.\ref{sum}) with respect to CHORUS and NOMAD and could
start data taking at the beginning of the next century.
Also proposals for a medium baseline search exist \cite{medium}. The present CERN \nue beam is coming
up to the surface again in a distance of about 17 km
away from the beam dump offering the chance for an experiment there.
\subsection{Long baseline experiments}
Several accelerators and underground laboratories around the world offer the possibility to
perform \lbls.
\smallskip\\
{\it KEK - \sk}:
The first of these experiments will be the KEK-E362 (K2K) experiment \cite{keksk} in Japan sending a \nue beam
from KEK to \sk. The distance is 235 km. A 1 kt front detector, about 1 km away from the beam dump will
serve as a reference and measure the \nue spectrum. The \nue beam with
an average energy of 1 GeV is produced by a 12 GeV proton beam dump. The detection method within \sk will be identical to that of their atmospheric \nue detection.
The beamline should be finished by the end of 1998 so the experiment could start data taking in 1999.
The experiment is of disappearance type. However an upgrade of KEK to a 50 GeV proton beam is
planned, which could
start producing data around 2004 and allow \ntau-appearance searches.
\smallskip\\ 
{\it Fermilab - Soudan}:
A big \nue program is also associated with the new Main Injector at Fermilab. The long baseline
project will
send a \nue beam to the Soudan mine about 735 km away from Fermilab. Here the MINOS experiment
\cite{minos} will be
installed. It consists of a front detector located at Fermilab close to COSMOS and a far detector at
Soudan. The far
detector will be made of 10 kt magnetized Fe toroids in 600 layers with 4 cm thickness interrupted by about 32000
m$^2$ active detector planes in form of streamer tubes with x and y readout to get the necessary tracking informations. The project could start
at the beginning of next century.
\smallskip\\
{\it CERN - Gran Sasso}:
A further program considered in Europe are \lbls using a \nue beam from CERN to Gran Sasso Laboratory.
The distance is about 732 km. Several experiments have been proposed for 
the \osz search. The first proposal is the ICARUS
experiment \cite{icarus} which will be installed in Gran Sasso anyway for the search of proton decay and 
solar neutrinos by using a liquid Ar TPC. A prototype  of 600 t is approved for installation in 1999.
A second proposal, the NOE experiment \cite{noe}, plans
to build a giant lead - scintillating fibre detector with a total mass of 4 kt. It will consist of 4 modules
followed by a module for muon
identification. A third proposal is a 27 kt water-RICH detector (AQUA-RICH) \cite{tom}, which could be installed either inside or outside the Gran
Sasso tunnel.
Finally there exists a proposal for a 1kt iron-emulsion sandwich detector (OPERA) \cite{niwa}.
It could consist of 2240 modules arranged in 140 planes each containing
4 $\times$ 4 modules. It is also under consideration for the Fermilab-Soudan beam.
\section{Summary and Conclusion}
Massive \nues allow a wide range of new phenomena in \nue physics, especially that
of \noszp Evidence for such \oszs comes from solar neutrinos, atmospheric \nues and the
\lsnd experiment. Terrestrial \nue experiments in form of nuclear reactors and 
high energy accelerators already exclude large parts of the parameter space because of non-observation of
\osz effects. Because 
the region of the MSW-solution for solar \nues are out of range for terrestrial experiments, current
and to a large extend future \osz experiments are motivated by the atmospheric \nue deficit, an eV-\nue as
dark matter candidate and 
a proof of the \lsnd results. 

\end{document}